\documentclass[12pt,preprint]{aastex}

\newcommand{\ml}{$M/L$}

\newcommand{\lsun}{L$_\sun$}

\newcommand{\om}{$\Omega_m$}

\newcommand{\sigv}{$\sigma_{v}$}
\newcommand{\sige}{$\sigma_{8}$}
\newcommand{\ngal}{$N_{gal}$}
\newcommand{\lam}{$\Lambda$}

\slugcomment{Last modified \today}

\begin{document}

\title{The Cluster Mass Function from Early SDSS Data:\\ 
Cosmological Implications}

\author{
Neta A. Bahcall\altaffilmark{1},
Feng Dong\altaffilmark{1},
Paul Bode\altaffilmark{1},
Rita Kim\altaffilmark{2},
James Annis\altaffilmark{3},
Timothy A. McKay\altaffilmark{4},
Sarah Hansen\altaffilmark{4},
Chris Miller\altaffilmark{6},
Josh Schroeder\altaffilmark{1},
James Gunn\altaffilmark{1},
Jeremiah P. Ostriker\altaffilmark{1},
Marc Postman\altaffilmark{5},
Robert C. Nichol\altaffilmark{6},
Tomotsugu Goto\altaffilmark{6},
Jon Brinkmann\altaffilmark{7},
Gillian R. Knapp\altaffilmark{1},
Don O. Lamb\altaffilmark{8},
Donald P. Schneider\altaffilmark{9},
Michael S. Vogeley\altaffilmark{10},
Donald G. York\altaffilmark{8}
}

\altaffiltext{1}{Princeton University Observatory, Princeton, NJ 08544}
\altaffiltext{2}{Department of Physics and Astronomy, The Johns Hopkins University, Baltimore, 
MD 21218}
\altaffiltext{3}{Fermi National Accelerator Laboratory, P.O. Box 500, Batavia, IL 60510}
\altaffiltext{4}{University of Michigan, Department of Physics, 500 East University, Ann Arbor, MI 48109}
\altaffiltext{5}{Space Telescope Science Institute, Baltimore, MD 21218}
\altaffiltext{6}{Department of Physics, Carnegie Mellon University, 5000 Forbes Avenue, Pittsburgh, PA 
15213-3890}
\altaffiltext{7}{Apache Point Observatory, 2001 Apache Point Road, P.O. Box 59, Sunspot, NM 88349-0059}
\altaffiltext{8}{The University of Chicago, Department of Astronomy and Astrophysics, 5460 S. Ellis Ave., 
Chicago, IL 60637}
\altaffiltext{9}{Department of Astronomy and Astrophysics, The Pennsylvania State University, University 
Park, PA 16802}
\altaffiltext{10}{Department of Physics, Drexel University, Philadelphia, PA 19104}

\begin{abstract}

The mass function of clusters of galaxies is determined from 400 deg$^2$ of early commissioning 
imaging data of the Sloan Digital Sky Survey; $\sim$300 clusters in 
the redshift range z = 0.1 - 0.2 are used. Clusters are selected using two independent 
selection methods: a Matched Filter and a red-sequence color magnitude technique. The two methods 
yield consistent results. The cluster mass function is compared with large-scale cosmological 
simulations. We find a best-fit cluster normalization relation of \sige \om$^{0.6}$ = 0.33 $\pm$ 0.03 
(for 0.1 $\la$ \om $\la$ 0.4), 
or equivalently \sige = ($\frac{0.16}{\Omega_m}$)$^{0.6}$. The amplitude of this relation is significantly 
lower than the previous canonical value, implying that either \om\  is lower than previously expected (\om 
= 0.16 if \sige = 1) or \sige \ is lower than expected (\sige = 0.7 if \om = 0.3) as suggested by recent 
results. The shape of the cluster mass function partially breaks this classic degeneracy; we find best-fit 
parameters of \om= 0.19 $\pm^{0.08}_{0.07}$ and \sige = 
0.9 $\pm^{0.3}_{0.2}$. High values of \om\ ($\ga$ 0.4) and low \sige \ ($\la$ 0.6) are excluded at $\ga$ 2$\sigma$. 

\end{abstract}

\keywords{
cosmology:observations--cosmology:theory--cosmological parameters--dark matter--galaxies:clusters:general--
large-scale structure of universe
}

\section{Introduction}

The abundance of clusters of galaxies as a function of mass places one of the strongest constraints 
on the amplitude of mass fluctuations on 8 $h^{-1}$ Mpc scale, \sige, and on the mass density 
parameter, \om. The present-day cluster mass function was the first observation
to suggest that the standard 
\om = 1 Cold Dark Matter (CDM) model has to be highly biased, with \sige $\sim$ 0.5 (i.e., a bias 
of $\sim$ 2, since the galaxy fluctuations amplitude is \sige$_{gal}$ $\sim$ 1), in order to match 
the observed cluster abundance. The cluster mass function also showed that low-density CDM models 
fit the cluster data well with little or no bias (i.e., \sige $\sim$ 1) \citep{bah92,whi93}. The 
mass function constraint, frequently called ``cluster normalization'' because of its powerful 
constraint on the linear mass power spectrum amplitude \sige, has provided the well known relation 
\sige \om$^{0.5}$ = 0.5 $\pm$ 0.05; this result was obtained from observations of both the cluster 
mass function \citep{bah92, bah93} and from the cluster temperature function \citep{edg90, hen91, 
whi93, kit96, eke96, VL96, eke98, pen98, mar98, hen00}. 
This relation implies that \om $\sim$ 0.3 if \sige $\sim$ 0.9 - 1; this latter 
value of \sige \ is suggested from other observations including 
cluster abundance evolution (\citealt{bah98, don00}, and references therein),
the flattening of the mass-to-light ratio on large scales \citep{bah95, bah00}, 
and the SDSS and 2dF large 
scale structure observations \citep{sza02, ver02}.
Similar \sige-\om \ normalization relations 
have been recently obtained from weak lensing observations on large scales 
(\citealt{van01, van02, hoe02, bac02, ref02}, and references therein).

More recently, using new X-ray cluster samples and different virial mass versus temperature 
relations (which are critical for a precise determination of the cosmological constraints), cluster 
normalizations that are either considerably lower (by $\sim$ 2$\sigma$;  \sige \om$^{0.5}\simeq$ 
0.4 $\pm$ 10$\%$ ) or higher ($\simeq$ 0.6 $\pm$ 10$\%$) than the above value have been reported 
(\citealt{bor01, ike02, sel02, rei02, via02, pie01};
here we converted all the relations 
to the same power-law slope of 0.5 for easier comparison).
An accurate 
determination of this parameter is important for two reasons. First, the normalization \sige$^{2}$ enters 
exponentially in the evolution of structure in the universe; a 20$\%$ change in \sige \  has a 
significant (exponential) effect on the evolution of structure with time and, of course, on the amount of bias 
in the universe (i.e., how mass traces light). Second, if we know \sige \ or \om \  from other 
observations, the above relation can be used to determine the second parameter. For example, if 
\sige $\sim$ 1, as suggested by some observations, then the implied value of \om \ differs by 
nearly a factor of two depending on whether the cluster normalization relation is \sige  
\om$^{0.5}$ = 0.6 $\pm$ 10$\%$ or 0.4 $\pm$ 10$\%$; these values imply \om = 0.36 $\pm$ 20$\%$ 
or 0.16 $\pm$ 20$\%$, respectively. 

Most of the previous analyses, which use the cluster temperature function, employ a smaller number 
of clusters, and assume a relation between virial 
cluster mass and temperature, which sensitively affects the results. The cosmological 
interpretations are generally based on comparisons with theoretical approximations such as the 
Press-Schechter formalism.

In this paper we use the early commissioning data from the Sloan Digital Sky Survey (SDSS: 
\citealt{yor00, sto02}) to determine a preliminary mass function of nearby clusters of galaxies 
and derive its cosmological constraints. The data cover about 400 deg$^{2}$ with $\sim$300 clusters at z = 
0.1 - 0.2 from each of two independent samples ($\sim$600 clusters in total) --- considerably larger 
than previous samples. The analysis does not use 
cluster virial masses, nor the virial mass temperature relation, which are more difficult to determine 
observationally. Rather, we use cluster masses observed within a fixed radius, as calibrated 
from the observed cluster luminosities and tested against cluster velocity dispersion. 
We use two independently selected cluster samples, identified by different algorithms --- the 
Matched Filter method and the color-magnitude maxBCG method; we find consistent results for the 
two samples. We compare the results directly with large scale (Gpc$^{3}$) cosmological simulations 
as well as with the Press-Schechter formalism to determine the cosmological constraints. 

Finally, we note that the current results are based on a very small fraction (4$\%$) of the ultimate SDSS 
10$^{4}$ deg$^{2}$ survey, which will yield thousands of clusters, many with velocity dispersions 
and weak gravitational lensing masses. The present mass function is therefore preliminary, 
intended to show the feasibility of using clusters from SDSS by utilizing the early commissioning 
data; larger and more accurate data will become available from SDSS in the near future.

\section{Cluster Selection from SDSS Commissioning Data}

The SDSS \citep{yor00} is a 5-band CCD imaging survey that will cover, when complete, 10$^4$ deg$^{2}$ 
of the high latitude North Galactic Cap, and a smaller deeper region in the South, followed by an 
extensive multi-fiber spectroscopic survey. The imaging survey is carried out in drift-scan mode in 
five SDSS filters, u, g, r, i, z, to a limiting point source magnitude of r$<$23 \citep{fuk96, gun98, 
lup01, hog01, smi02, pie02}. The 
spectroscopic survey will target nearly one million galaxies to approximately r$<$17.7, with a median 
redshift of z$\sim$0.1 \citep{str02}, and a smaller deeper sample of $\sim$ 10$^{5}$ Luminous Red 
Galaxies to r$\sim$19 and z$\sim$0.5 \citep{eis01}.

In this paper we use 379 deg$^{2}$ of the early commissioning data of SDSS imaging, covering the area 
$\alpha$(2000) = 355.0\degr to 56.0\degr, $\delta$(2000) = -1.25\degr to 1.25\degr; and $\alpha$(2000) = 
145.3\degr to 236.0\degr, $\delta$(2000)= -1.25\degr to 1.25\degr\ (runs 94/125 and 752/756; \citealt{sto02}). 
Clusters of galaxies were 
selected from these imaging data using, among others, a Matched-Filter method \citep{kim02a, kim02b} 
and an independent color-magnitude maximum-likelihood Brightest Cluster Galaxy method (maxBCG; 
\citealt{ann02}). These methods are briefly described below. A detailed comparison between these 
independent cluster selection methods and their properties is given in \citet{bah02}. Here we use 
clusters selected from these techniques to determine a preliminary mass function of nearby clusters 
of galaxies. 

The Matched Filter method HMF (Hybrid Matched Filter; \citealt{kim02a}) is a hybrid of the Matched Filter 
\citep{pos96} and the Adaptive Matched Filter techniques \citep{kep99}.  This 
method identifies clusters in imaging data by finding peaks in a cluster likelihood map generated 
by convolving the galaxy survey with a filter based on a model of the cluster and field galaxy 
distribution. The cluster filter is composed of a projected density profile model for the galaxy 
distribution (Plummer law profile), and a luminosity function filter (Schechter function), using 
the typical parameters observed for galaxy clusters (within a radius of 1 $h^{-1}$ Mpc). The 
HMF method identifies the highest likelihood clusters in the imaging data and determines their 
best-fit estimated redshift (z$_{est}$) and richness (\lam); the best-fit richness is 
proportional to the total cluster maximum likelihood luminosity within a radius of 1 $h^{-1}$ Mpc. A relatively high 
threshold has been applied to the HMF cluster selection ($\sigma>$5.2, \citealt{kim02a}); therefore, 
the selected clusters have typical richness of \lam $>$ 20-30 (i.e., L$_{cl}$($<$ 1$h^{-1}$ Mpc) 
$>$ 20L$^{*}\sim$2$\times$10$^{11}$ $h^{-2}$ \lsun ). This threshold corresponds to clusters poorer 
than Abell richness class 0. (For more details see \citealt{kim02a}).

The maxBCG method \citep{ann02} is based on the fact that the brightest cluster galaxy (BCG) 
generally lies in a narrowly defined space in luminosity and color (see, e.g, \citealt{hoe85, gla00}). For 
each SDSS galaxy, a BCG likelihood is calculated as a function of redshift based on the galaxy 
color ($g$-$r$ and $r$-$i$) and magnitude. The cluster likelihood is then weighted by the number of nearby red galaxies 
(located within 1 $h^{-1}$ Mpc projected separation) that are within the color-magnitude region 
expected for the relevant cluster E/S0 galaxy ridgeline. This 
combined likelihood is used for cluster identification. The likelihood is calculated as a function 
of redshift from z = 0 to 0.5, at 0.01 intervals. The best estimated redshift is that which maximizes 
the cluster likelihood. Since elliptical galaxies possess very regular colors, they provide excellent 
photometric redshift estimates for their parent clusters. The richness estimator, \ngal, is defined as 
the number of red E/S0 ridgeline member galaxies that are brighter than M$_{i}$ = -20.25 (i.e., 1 mag 
fainter than L$^{*}$; $h$ = 1), and are located within a 1 $h^{-1}$ Mpc radius of the BCG. (For more 
details see \citealt{ann02}).

The HMF cluster catalog contains clusters with richness \lam $>$ 20 and redshift z$_{est}$ 
$<$ 0.5 \citep{kim02a, kim02b}. The selection function for this sample has been determined using 
simulated clusters (see above references). The HMF redshift uncertainty is determined to be $\sigma
_{z}$ = 0.03 (by comparison with measured redshifts, \citealt{bah02}); the redshift uncertainty 
of the maxBCG clusters is $\sigma_{z}$ = 0.02. 

In this paper we determine the abundance of HMF clusters as a function of richness for nearby 
clusters (z = 0.1 - 0.2) and use the observed richness - mass relation to determine a preliminary 
mass function for the HMF clusters. A similar analysis is carried out for the independently selected 
maxBCG clusters and the results compared. Each of the independent samples contains $\sim$300 clusters 
within the redshift (z = 0.1 - 0.2) and richness (\lam $\geq$ 30 and \ngal$\geq$ 10) limits used in this analysis.

\section{The Cluster Mass Function}

\subsection{HMF Cluster Mass Function}

We determine the mass function of nearby clusters of galaxies using HMF clusters with richness \lam 
$\geq$ 30 and redshift z = 0.1 - 0.2. (At z $<$ 0.1, the number of clusters is small and their selection 
less effective; we thus restrict our analysis to the above range.) To minimize false-positive 
detections we use the VC1 sample (Visually Confirmed sample, \citealt{kim02a, kim02b}) which contains 
$>$ 80$\%$ of all \lam $\geq$ 30 HMF clusters, increasing to $>$ 90$\%$ for \lam $\geq$ 50 clusters. 
The total number of VC1 clusters observed within this redshift and richness range is 294 (uncorrected for 
selection function). Each cluster is 
corrected by the appropriate selection function for the given cluster richness and redshift as 
determined from cluster simulations \citep{kim02a}. The cluster abundance as a function of 
richness, from \lam $\geq$ 30 to \lam $\geq$ 70, is obtained by dividing the above volume-limited corrected
cluster count by the relevant volume (z = 0.1 - 0.2). A flat \om = 0.3 cosmology is assumed for the volume 
calculation, and a Hubble 
constant of H$_{0}$ = 100 $h$ km/s/Mpc is used. (When fitting to different cosmologies in Section 4, 
the proper self-consistent cosmological volume is used for each \om \ value.)

Two corrections are applied to the cumulative cluster richness function. First, we correct the 
abundance of clusters above a given richness, $n$ ($\ge$ \lam), for the effect of redshift uncertainty 
in the HMF clusters, $\sigma_{z}$ =
0.03 (see Section 2). The correction factor is determined using Monte Carlo simulations of realistic 
cluster distribution with redshift and richness, which is convolved with the 
observed Gaussian scatter in redshift, $\sigma_{z}$ = 0.03. We find that the redshift uncertainty 
has a small effect, causing about 10$\%$ more clusters to be scattered into the z = 0.1 - 0.2 volume than 
are scattered out. 
We correct the cluster abundances downward by this small correction. Second, we correct the derived 
cluster abundance for the effect of uncertainty in the HMF richness, estimated to be 20$\%$ based 
on cluster simulations. We use Monte Carlo simulations with a realistic richness function, 
convolve it with the known observational selection function to produce the observed number of 
clusters as a function of true richness and then scatter the richness with the observed uncertainty 
to yield the observed richness function. Comparing the observed and true richness functions in 10$^3$ 
simulations we determine the proper correction factors and their dispersion, which we apply to the data. We 
find that the observed abundances are larger than the true ones, as expected due to the excess scatter of the 
more numerous low richness clusters to higher richness; this effect is 10$\%$ at \lam 
$\sim$ 30 - 40, increasing to 35$\%$ - 55$\%$ at \lam $\sim$ 60 - 70. We correct the cluster abundances 
for this effect, and use the observed variance in the final error analysis discussed below. 

The uncertainties in the observed cluster abundance include the statistical uncertainties (N$^
{\frac{1}{2}}$), the uncertainties in the selection function ($\pm$15$\%$) and in the false-positive 
correction ($\pm$15$\%$), and the uncertainties derived from the Monte Carlo simulations for each 
of the two corrections above (the redshift correction factor has an uncertainty of 4$\%$ to 42$\%$ 
for the range \lam $\geq$30 to \lam $\geq$70, and the richness correction factor has an uncertainty of 
3$\%$ to 23$\%$ for the same range). 

To determine a cluster mass function from the above cluster richness function we need to convert 
the cluster richness thresholds to a mean cluster mass. Throughout this paper we use cluster mass 
within a given fixed radius (not virial mass); this mass is more accurately obtained from 
observations since the virial radius is not precisely known. We convert richness to mass in two 
independent ways, both from observations. First, we use the mean cluster luminosity measured in the SDSS 
data for all clusters stacked as a function of their richness. The cluster luminosity is observed within a radius of 0.6  
$h^{-1}$ Mpc, in the r-band, for galaxies brighter than M$_{r}$ = -19.8 (K-corrected for each galaxy type 
following \citealt{fuk96}), and corrected for a similarly determined local 
background in five separate locations (which allows us to account for the variance in the background correction; 
\citealt{han02, bah02}). We use the mean observed cluster luminosity L$_{0.6}$ for clusters with richness 
threshold of \lam=30, 40, 45, 50, 60 and 70.  
The observed mean luminosity L$_{0.6}$ of the stacked clusters is presented as a function of 
richness in Figure 1. We note that any biases or uncertainties in the 
richness parameter (e.g., \citealt{kim02a}) are calibrated out in this procedure since the actual mean cluster 
luminosities are directly measured by this method. The richness parameter serves only 
as a tracer; a richness bias will properly calibrate itself by the measured mean 
luminosity (as is in fact seen by the non-linear relation between $L_{0.6}$ and \lam). 
The cluster luminosity is corrected to include the unobserved 
faint-end of the cluster luminosity function, that is, all galaxies fainter than M$_{r}$ = -19.8. For 
the observed SDSS Schechter luminosity function parameters of the HMF clusters (within 0.6 $h^{-1}$ Mpc), 
$\alpha$ = -1.08 $\pm$ 0.01 and M$^{*}_{r}$ = - 21.1 $\pm$ 0.02 (\citealt{han02}; see also \citealt{got02}; 
$h$ = 1), we adopt a correction factor of 1.42 $\pm$ 0.08 for the added 
contribution of faint galaxies to the total HMF cluster luminosity. The cluster mean luminosity is then converted 
to cluster mass, M($<$ 0.6 $h^{-1}$ Mpc physical), using the mean observed cluster $M/L_{r}$ ratio for each 
richness threshold \citep{bahc02}. The observed best-fit \ml\ is used (based on the means of 20 clusters and 33 
groups): $M/L_{v,tot}$ (z=0) = 142$\pm$32 + (23$\pm$5) $T_{kev} \ h$ \citep{bahc02}. The mild increase of \ml\ with 
temperature, $T_{kev}$ (seen both in observations and in simulations, e.g., \citealt{bah00}), is accounted for at each 
richness threshold using the observed correlation between richness and velocity dispersion (see below) and the 
observed mean relation between velocity dispersion and temparature (\sigv = 332 $T_{kev}^{0.6} km s^{-1}$, \citealt{lub93}). 
This effect is small for the range of cluster temperatures studied here ($T\simeq$ 0.9 to 4 kev).
The mean observed $M/L_{v,tot}$ at z=0 is converted to $M/L_{r}$ (where $L_r$ is the relevant SDSS 
Petrosian $r$ luminosity) using the conversions given by \citet{fuk96}, \citet{bahc02}, \citet{str02}. We use 
$L_r$ = 0.85 $L_{r,tot}$ (\citealt{str02}, for $\sim$ 60 - 70$\%$ of cluster light 
contributed by early type galaxies), $M/L_{r,tot}$ = 0.94 $M/L_{v,tot}$, 
and $M/L_r$ (z = 0.17) = 0.943 $M/L_r$ (z = 0) \citep{car96, bahc02}. The 
above yields $M/L_r$ values (at z=0.17) that range from 170 at \lam$\geq$30 to 235 at \lam$\geq$70. 
The mean mass of clusters (within 0.6 $h^{-1}$ Mpc) is then determined for the relevant richness 
thresholds (\lam $\geq$30 to \lam $\geq$70). 

The uncertainties in the mean mass estimates are derived from the combined uncertainties in the 
observed mean luminosity-richness relation, the uncertainty in the observed mean \ml\ ratio, and 
the smaller uncertainties in the corrections applied above. In order to determine the proper 
uncertainties in the luminosity-richness relation that are relevant for the mean cluster mass 
estimates, we generate 10$^3$ Monte Carlo simulations with a realistic L-\lam\ relation 
and richness function. We introduce a Gaussian redshift scatter of $\sigma_{z}$ 
= 0.03 as well as a $\sim$ 20$\%$ to 30$\%$ uncertainty in individual cluster luminosities. 
We recover the mean "observed" L-\lam\ relation from the 10$^3$ simulations and the mean luminosities 
at the relevant cluster richness thresholds (for the observed redshift range). The recovered mean relation is 
consistent with the input L-\lam\ relation. The derived 1$\sigma$ variance 
in the recovered L-\lam\ relation from the simulations ranges from 11$\%$ at \lam=30 to 25$\%$ at 
\lam=70. We use these uncertainties in estimating cluster mass uncertainties. The uncertainty in the 
mean observed \ml \ ratio, 15$\%$ for the relevant cluster richnesses, is combined with an 
additional 8$\%$ uncertainty in the conversion factors described above and a 6$\%$ uncertainty in the 
luminosity function faint end extrapolation (see above). The mass 
uncertainties thus range from 20$\%$ at \lam $\sim$30 to 31$\%$ at \lam $\sim$70. 
The cluster abundances have been corrected for this scatter using Monte Carlo simulations, as described above 
for the richness-function abundance correction.

For comparison with other commonly used cluster masses, as well as for direct comparison with available 
cosmological simulations, we also determine the mass function for cluster masses within two additional frequently 
used radii: the slightly smaller radius of 0.5 $h^{-1}$ Mpc, using the observed mean luminosities L$_{0.5}$, 
and, for illustration purposes, also the larger comoving radius of 1.5 $h^{-1}$ Mpc; the latter is obtained by 
extrapolating the 0.6 $h^{-1}$ Mpc luminosity to 1.5 $h^{-1}$ Mpc comoving radius (= 1.28 
$h^{-1}$ Mpc at z = 0.17) using the typical observed luminosity profile in clusters ( $\rho$$_{L}$ 
$\sim$ R$^{-2}$ for R $<$ R$_{200}$ and $\sim$ R$^{-2.4}$ for R $\ga$ R$_{200}$, where R$_{200}$ is 
the radius within which the cluster overdensity is 200 times the critical density; \citealt{car97, fis97}). 

These two mass functions are compared in Figure 2 with the mass function obtained from 
large--scale cosmological simulation \citep{bod01} of the concordance LCDM model \citep{bah99}: 
\om= 0.3, \lam= 0.7, \sige= 0.9 (i.e. \sige \om$^{0.5}$ = 0.49), and $h$= 0.67.
This simulation used a 1 $h^{-1}$ Gpc box size and 1024$^3$ dark matter particles, 
with a particle mass of 2.3 $\times$ 10$^{10}$ $h^{-1}$ M$_\sun$, and softening length of 
14 $h^{-1}$ Kpc \citep{bod01}.  Such a large box ensures a statistically valid sample of simulated clusters,
and the high particle number ensures that the clusters are well resolved --- 
for the smallest clusters considered here there are over $10^3$ particles within 0.5 $h^{-1}$ Mpc.
The details of the simulation and the method of computing the mass function are
described in \citet{bod01}. The simulated mass function is presented as a function of M($<$ 0.5 
$h^{-1}$ Mpc physical) and M($<$ 1.5 $h^{-1}$ Mpc comoving) at z = 0.17, for direct comparison 
with the observations. Figure 2 shows that the shape of the SDSS mass function agrees well with that 
expected from the cosmological simulations but the normalization is significantly lower than expected 
from the concordance model. The best-fit function, with a lower \sige-\om \ amplitude, is also presented in 
Figure 2; it is discussed in Section 4.

The observed HMF cluster mass function for M($<$0.6 $h^{-1}$ Mpc) is presented in Figure 3. As a further 
consistency test, we estimate mean cluster masses using an entirely independent method: the 
observed correlation between mean cluster richness and cluster velocity dispersion. We use cluster velocity 
dispersions of 19 clusters determined from the SDSS spectroscopic survey (for clusters with $\sim$30 to 160 redshifts) 
as well as from several Abell clusters available in the literature (\citealt{maz96, sli98}; Abell 168, 295, 957, 
1238, 1367, 2644). Even though the number of clusters with measured velocity dispersion is not large and the 
scatter considerable, a clear correlation between median velocity dispersion and richness is observed, as expected; 
we find a best-fit relation \sigv(km s$^{-1}$) $\simeq$ 10.2 \lam. We estimate mean cluster mass 
(within 0.6 $h^{-1}$ Mpc) from this relation and use it to illustrate consistency with the mass function determined 
from the entirely independent cluster luminosity method discussed above. We use the observed relation between cluster 
mass and cluster velocity dispersion derived from observations of weak gravitational lensing of 
clusters: M($<$ 0.6 $h^{-1}$ Mpc) = 0.0717 $k_{\delta}$ $\sigma_{100}^{1.67}$ 10$^{14}$ $h^{-1}$ M$_\sun$ 
(where $\sigma_{100}$ is in 100 km s$^{-1}$; \citealt{hjo98}; also \citealt{bahn02}). This relation is obtained 
from the observed relations $M/R$ = 0.88 $k_{\delta}$ T(kev) for R $<$ 1 $h^{-1}$ Mpc, where $k_{\delta}$ is the small 
overdensity correction factor ($k_{\delta}$ = 0.76, 0.9, 1, 1.1, 1.15, respectively, for cluster overdensity 
of $\delta$ = 100, 250, 500, 1000, 2500; see references above and \citealt{evr98}), and \sigv($km s^{-1}$) = 332 
$T_{kev}^{0.6}$ \citep{lub93}. The cluster mass function determined from this independent method, performed as a 
consistency check, is in full agreement with the mass 
function determined earlier using cluster luminosities; the results are compared in Figure 3. The velocity 
dispersion comparisons from the two methods --- i.e., the velocities inferred from the cluster luminosity-mass 
method and the directly observed velocity dispersions are shown as a function of richness in 
Figure 4. The excellent agreement between these two independent methods supports the mass determination 
discussed above. 

\subsection{maxBCG Cluster Mass Function}

For comparison, we also determine the cluster mass function from the independently selected 
maxBCG clusters. This method uses a completely independent selection criterion: the maxBCG 
selection technique assumes no cluster filters or profiles; rather, it selects clusters based on 
the red colors and magnitudes of the brightest cluster galaxies (Section 2). A comparison of the 
two mass functions can therefore provide further support for the above results.

We follow the same procedure for the maxBCG clusters as described above
for the HMF clusters. We use the observed mean luminosity L$_{0.6}$ of all stacked maxBCG clusters as a function of 
richness, \ngal\ (where \ngal \ is the maxBCG cluster richness, Section 2); the data are presented in Figure 1. 
We extrapolate the luminosity to the faint-end of the 
cluster luminosity function (within 0.6 $h^{-1}$ Mpc; $\alpha$ = -1.05 $\pm$ 0.01, M$^{*}_{r}$ = 
-21.25 $\pm$ 0.02 for the maxBCG clusters for 
$h$ = 1, \citealt{han02}; see also \citealt{got02}), yielding a correction factor of 1.34 $\pm$ 0.06, 
and convert the cluster luminosity to mean cluster mass using the 
mean observed \ml \ ratios. All maxBCG clusters (357 clusters) with richness \ngal $\geq$ 10 
(comparable in richness to HMF clusters with richness \lam $\ga$ 30; \citealt{bah02}) in the 
redshift range z = 0.1 - 0.2 are used. Corrections and uncertainties are calculated as described above 
(with $\sigma_{z}$= 0.02, $\Delta$\ngal = 10$\%$ - 15$\%$). The selection function and the false-positive 
correction factor for the \ngal$\geq$10 maxBCG clusters at z = 0.1 - 0.2 are estimated from simulations 
to be $\sim$ 0.9 - 1 $\pm$ 15$\%$ each. The fraction of HMF clusters that are found by the maxBCG method is 61$\%$ 
(for maxBCG matches with \ngal$\geq$6 located within 1 $h^{-1}$ Mpc projected separation). This is consistent with 
the maxBCG selection function ($\sim$85$\%$ for \ngal$\geq$6), the HMF false-positive detection rate ($\sim$20$\%$, 
for \lam$\geq$30), and the smaller correction due to redshift uncertainty (see \citealt{bah02}). The overlap rate 
decreases considerably if only \ngal$\geq$10 maxBCG matches are considered for the HMF clusters (within 1 $h^{-1}$ 
separation and $\pm$0.05 in redshift). This is as expected due to the large scatter of clusters across the richness 
threshold; Monte Carlo simulations indicate that a richness cut reduces the overlap by $\sim$ 55$\%$ - 60$\%$ 
\citep{bah02}.

There are 58 Abell clusters located within the current survey area; all 58 clusters are detected by the combined 
HMF and maxBCG clusters (maxBCG finds all 58 clusters; HMF finds 49 of the 58 clusters, consistent with the respective 
selection functions). A few of the clusters are detected below the thresholds used here (i.e., \lam$<$30, 
\ngal$<$10, z$<$0.1, or z$>$0.2). In addition, there are 5 X-ray clusters from the XBACs sample \citep{ebe96} 
located within the survey region. All 5 are detected by the HMF and maxBCG methods.

The maxBCG cluster mass function is presented in Figure 5; it is 
superposed, for comparison, on the HMF mass function from Figure 3. A good agreement 
between the two independent mass functions is observed. This agreement provides further support to the 
above estimate of the SDSS cluster mass function. In addition, we use the best-fit relation observed 
between mean cluster velocity dispersion and richness for maxBCG clusters (21 clusters with measured dispersions) 
as an additional test; the observed median relation, \sigv(km/s) = 93 \ngal$^{0.56}$ (Figure 4), is used to independently 
estimate cluster masses using M ($<$ 0.6 $h^{-1}$ Mpc) = 0.0717 $k_{\delta}$ $\sigma_{100}^{1.67}$ 10$^{14}$ 
$h^{-1}$ M$_\sun$, as described above. The results of the two methods (Figure 5) are consistent with 
each other. 
	
We can compare the observed cluster mass function with the recently observed cluster temperature 
function by \citet{ike02}. For this purpose we use the observed relation discussed above, M ($<$ 0.6 
$h^{-1}$ Mpc) = 0.53 $k_{\delta}$ T(kev) 10$^{14}$ $h^{-1}$ M$_\sun$, to derive an approximate temperature 
function from the above mass function. We find 
a good agreement between the mass function (based on 300 clusters at z = 0.1 - 0.2) and the Ikebe et al.
temperature function (based on 60 X-ray clusters, mostly at z $<$ 0.1) (accounting for the slightly higher redshift of 
the SDSS sample, in accord with cosmological simulations; \citealt{bod01}). 
The SDSS mass function reaches to poorer clusters, of lower temperature (T $\sim$ 1 kev), 
as compared with the X-ray temperature function (T $>$ 2 kev); because of the small area covered, the 
current SDSS sample does not contain the most massive clusters--- these highest mass clusters will become 
available as the sample size increases. The agreement between these independent 
determinations provides further support of the current cluster mass function results. 

\section{Cosmological Implications}

The cluster mass function places one of the most powerful constraints on the cosmological 
parameters \om \  and \sige; it determines the important cluster normalization relation, i.e., 
the value of \sige \ as a function of \om.

Early data of the cluster mass function \citep{bah92}, and the cluster temperature function 
\citep{hen91, whi93, eke96, eke98, pen98}, provided a cluster normalization relation of 
\sige \om$^{0.5}$ $\simeq$ 0.5 ($\pm$ $\sim$10$\%$; see above references for details). This 
powerful relation implies that for \sige  $\sim$ 1, \om $\sim$ 0.25 (with slight differences 
depending on a flat versus open cosmology). For \om = 1, the required normalization of \sige  
= 0.5 implies a strong bias which is not supported by observations \citep{bah95, bah00, bah98, fel01, 
ver02, lah02}.

More recently, using different X-ray cluster samples and different relations between virial 
mass and cluster temperature, a somewhat lower normalization value has been suggested \citep{bor01, ike02, 
sel02, rei02, via02}, although higher values have also been reported \citep{pie01}. 
In this section we compare the
preliminary mass function of SDSS clusters
with analytic predictions to
determine the best-fit cosmological parameters.

The mass function for a given cosmology can be predicted
using the analytic formalism of \citet{PS74}, as in
for example \citet{eke96, kit96, VL96, hen00}; and \citet{rei02}.
While fairly successful in matching the results of
N-body simulations, the standard P-S formalism tends to 
predict too many low mass clusters and too few higher mass clusters.
An improved fitting formula which better reproduces the
results of N-body simulations is given by
\citet{JFWCCEY01}; we will use this in preference to
the standard P-S formula.

For a given set of cosmological parameters, we begin by
calculating the linear matter power spectrum using the publicly 
available CMBFAST code \citep{CMBFAST96}.   Knowing the power 
spectrum, we calculate the variance of the linear density 
field and thus find the mass function with equation B3 of
\citet{JFWCCEY01}. 
As \citet{HuKrav02} show,
this formula is appropriate for a definition of cluster mass 
within a sphere enclosing a mean overdensity 
which is a fixed multiple of the mean density.  
Because the HMF and maxBCG masses are instead
derived within a fixed radius (independent of density or mass),
we adjust the analytic masses using the mass distribution 
corresponding to the NFW density profile \citep{NFW97},
which provides an accurate representation of N-body results.
To do this we follow the method presented in the Appendix
of \citet{HuKrav02}.

The resulting analytic prediction, using cluster mass within two different
radii of 0.5 $h^{-1}$ Mpc (physical) and 1.5 $h^{-1}$ Mpc (comoving), is shown for 
the concordance model in Figure 2 as a dotted line.
This analytic function is in excellent agreement with the direct N-body simulation (dashed 
line) over the relevant range of masses (though at higher or lower masses 
the agreement may not be so close).

To determine the best-fit mass function and the implied values of $\Omega_m$ and $\sigma_8$,
we compare the differential binned mass function to the above theoretical prediction using a standard $\chi^2$ 
procedure. The last data point in each sample is not included in the best $\chi^2$ fit 
determinations since this point contains only a few clusters ($\sim$7) with considerably less well determined 
mass. Only spatially flat models are considered. The Hubble constant is 
kept fixed at H$_0$ = 72 km s$^{-1}$ Mpc$^{-1}$ \citep{FMGetal01}, with a baryon density of $\Omega_b h^2=0.02$ 
\citep{BNT01}, and CMB temperature $T_{CMB}=2.726$ \citep{Mather94}. We assume a primordial power 
spectral index $n=1$.

The results, presented in Figures 2, 3 and 5, show a good agreement between the shape of 
the observed and theoretical LCDM mass function, but the observed function has a  significantly lower 
normalization than the canonical value of \sige \om$^{0.5}$ =
0.5 (the latter indicated by the dashed and dotted
curves in Figure 1, representing the concordance LCDM model 
with \om = 0.3 and \sige  = 0.9). Model simulations with a somewhat 
lower value of \sige  \om$^{0.5}$ $\simeq$ 0.45 (for a flat Quintessence model and Open 
CDM, see \citealt{bod01}; not shown here to avoid crowding) also exhibit higher normalization 
than observed.

The best-fit mass function is presented in Figures 2 and 3 for the HMF clusters, and in Figure 5 for both 
the maxBCG and HMF clusters; the two independent best fit functions yield similar results (Figure 5). 
The cosmological constraints derived from the $\chi^2$ minimization
are summarized in Figure 
6, showing the allowed parameter range of \om -\sige \  for both the HMF and the maxBCG samples. 
The best-fit parameters are \om = 0.175 $\pm^{0.08}_{0.07}$, \sige = 0.92 $\pm^{0.25}_{0.20}$ (1-$\sigma$) 
for the HMF clusters, 
and \om = 0.195 $\pm^{0.09}_{0.07}$, \sige = 0.9 $\pm^{0.3}_{0.2}$ for the maxBCG clusters (Figures 2, 3 and 5). 
The best-fit contours in Figure 6 show that high values of 
\om ($>$ 0.4) and low values of \sige ($<$ 0.6) are ruled out by the data at $\ga$ 2-$\sigma$; these yield 
mass functions that are too steep compared to the data. On the 
other hand, low values of \om (down to $\sim$ 0.1) and high values of \sige (up to $\sim$ 1.2) are 
supported by the data.
These results are  obtained using the  M($<$ 0.6 $h^{-1}$ Mpc) mass functions;
similar results are obtained with M($<$ 1.5$h^{-1}$ Mpc comoving).

The mean best-fit parameters of the observed mass function (with 1-$\sigma$) are:
\begin{equation}
       \Omega_m = 0.19 \pm^{0.08}_{0.07}  
\end{equation}
\begin{equation}
       \sigma_8 = 0.9 \pm^{0.3}_{0.2}
\end{equation}

The best-fit normalization relation (Figure 6) is:
\begin{equation}
       \sigma_8\Omega_m^{0.6} = 0.33 \pm0.03 \ \ \ \ \ \ \ \ \  (0.1\la\Omega_m\la0.4)  
\end{equation}
or, equivalently, 
\begin{eqnarray}
       \sigma_8 = (\frac{0.16}{\Omega_m})^{0.6} \pm10\%\ \ \ \ \ {\rm .}
\end{eqnarray}

For comparison with previous results, this relation (3) is 20$\%$ lower than the standard normalization 
value of \sige \om$^{0.5}$ = 0.5 (or similarly, \sige \om$^{0.6}\simeq$ 0.44). As discussed earlier, 
this conclusion has a non-negligible implication for 
\sige \  and \om \ as seen in equations (1) and (2) for the best-fit parameters (and discussed below).


The results are consistent with the recent temperature function results of \citet{ike02} and \citet
{sel02}. The higher amplitude obtained by some of the earlier work is most likely due to a combination of factors including 
use of the uncertain and sensitive theoretical (rather than observational) mass-temperature relation 
(when applied to X-ray clusters), smaller sample size, and overestimated cluster abundance (in some optical samples). 
Recent weak lensing observations on large scales yield results that range from \sige 
 \om$^{0.5}$ $\sim$ 0.4 to 0.6 (or \sige \om$^{0.6}$ $\sim$ 0.34 to 0.52 when converted to a power law slope of 0.6, 
at \om $\sim$ 0.25, for easier comparison; see references in Section 1); the low end of this range is consistent with the 
current results, but the high normalization values reported are inconsistent at the 2-$\sigma$ level. 
A complementary analysis of the maxBCG cluster halo occupation function (Annis et al., in preparation) yields 
consistent results with those obtained above.

Our best-fit \sige-\om \ constraints are compared with previous results in Figure 7. This comparison illustrates the 
agreement of recent temperature function results with the current constraints, and shows the wide range among the 
earlier, higher \sige(\om) normalization results. The weak lensing analyses yield constraints that lie mostly at the higher 
\sige(\om) range; the lensing results of \citet{hoe02} are consistent with our current constraints.

This new cluster normalization has important implications for \om. It is frequently assumed that 
\om\  is 0.3, and the above relation is used to determine \sige\  (thus typically referred to as 
"low normalization", implying a lower than expected \sige \  value for \om = 0.3). However, the value of \om \ 
is not accurately known, and could be as low as $\sim$ 0.15 (see, e.g., \citealt{bah00}). At the same time, 
there are several measurements that suggest that \sige $\simeq$ 0.9 - 1. This ``high'' normalization 
is obtained from the very mild evolution observed in the cluster abundance to z$\sim$1 \citep{bah98, 
don00}; the flattening of the \ml \ function on large scales \citep{bah95, bah00};  SDSS observations 
of the galaxy power spectrum on large scales \citep{sza02}; and the observation of no-bias in 
the galaxy distribution in the 2dF and infrared surveys (\citealt{ver02, fel01}; but see, however, 
\citealt{lah02}). This observationally suggested normalization of \sige $\simeq$ 0.9 - 1 is fully consistent with 
our best-fit value for the SDSS cluster mass function (equations 1 and 2); it implies a low mass density of 
\om $\simeq$ 0.19. This is also consistent with the low \om \ value indicated by the \ml \ function on large scales.

\section{Conclusions}

We determine the mass function of nearby clusters of galaxies using $\sim$ 300 clusters at z = 0.1 - 0.2 
selected from $\sim$ 400 deg$^2$ of early SDSS commissioning data. Two independent cluster samples 
are used based on the Matched Filter and the color-magnitude maxBCG methods. The two 
samples yield consistent results. The analysis uses cluster masses within a fixed radius. 
The mass function is compared directly with large, Gpc$^3$ cosmological 
simulations. We find a best-fit cluster normalization relation of \sige \om$^{0.6}$ = 0.33 $\pm$ 
0.03, or equivalently \sige = ($\frac{0.16}{\Omega_m}$)$^{0.6} \pm$ 10$\%$. This result is significantly 
lower than the previous canonical value of \sige \om$^{0.5}$ = 0.5 $\pm$ 0.05. The shape of the cluster mass function 
partially breaks this degeneracy; we find best-fit parameters of \om = 0.19 $\pm^{0.08}_{0.07}$ and 
\sige = 0.9 $\pm^{0.3}_{0.2}$. These values are consistent with the independent observationally suggested 
normalization of \sige $\simeq$ 0.9 - 1 observed from cluster abundance evolution \citep{bah98, don00}, 
the flattening of the \ml \ function on large scales \citep{bah00}, and SDSS, 2dF, and infrared large-scale 
structure observations \citep{sza02,ver02,fel01}.

These preliminary results from early commissioning data of 4$\%$ of the ultimate SDSS survey will 
be greatly improved as additional SDSS data become available for thousands of clusters, many with 
measured velocity dispersions and weak lensing masses. Improvements in the sample size and accuracy, and in the relevant 
scaling relations are needed in order to achieve greater precision in the determination of the mass function and its 
cosmological implications.

The Sloan Digital Sky Survey is a joint project of The University of Chicago, Fermilab, the Institute for 
Advanced Study, 
the Japan Participation Group, The Johns Hopkins University, Los Alamos National Laboratory, the 
Max-Planck-Institute for Astronomy (MPIA), the Max-Planck-Institute for Astrophysics (MPA), New Mexico 
State University, Princeton University, the United States Naval Observatory, and the University of
Washington. Apache Point Observatory, site of the SDSS telescopes, is operated by the Astrophysical 
Research Consortium (ARC). Funding for the project has been 
provided by the Alfred P. Sloan Foundation, the SDSS member institutions, the National Aeronautics and Space 
Administration, the National Science Foundation, the Department of Energy, the Japanese Monbukagakusho, 
and the Max Planck Society. The SDSS Web site is http://www.sdss.org.

\clearpage


\epsscale{1.25}
\begin{figure}
\plottwo{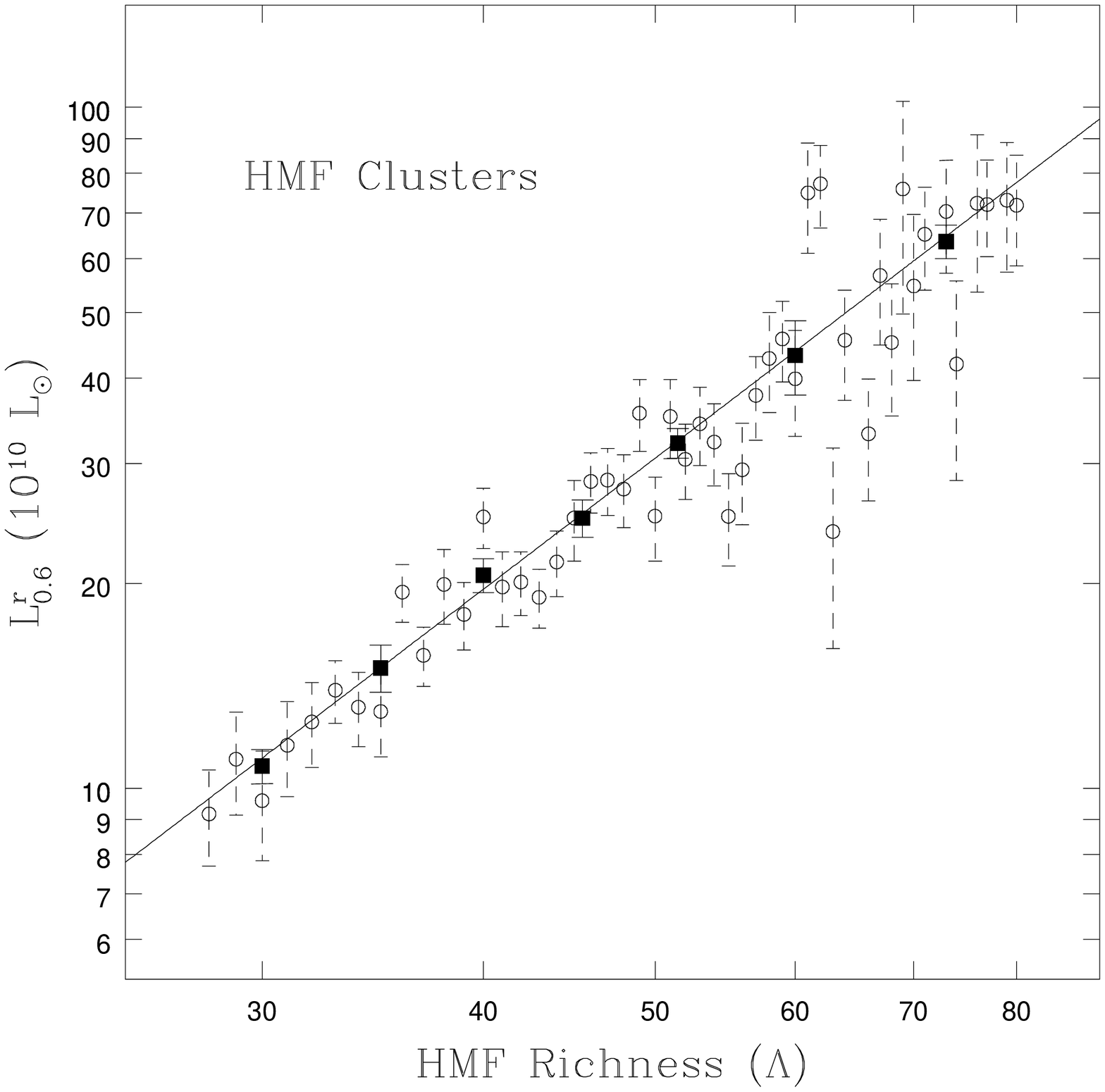}{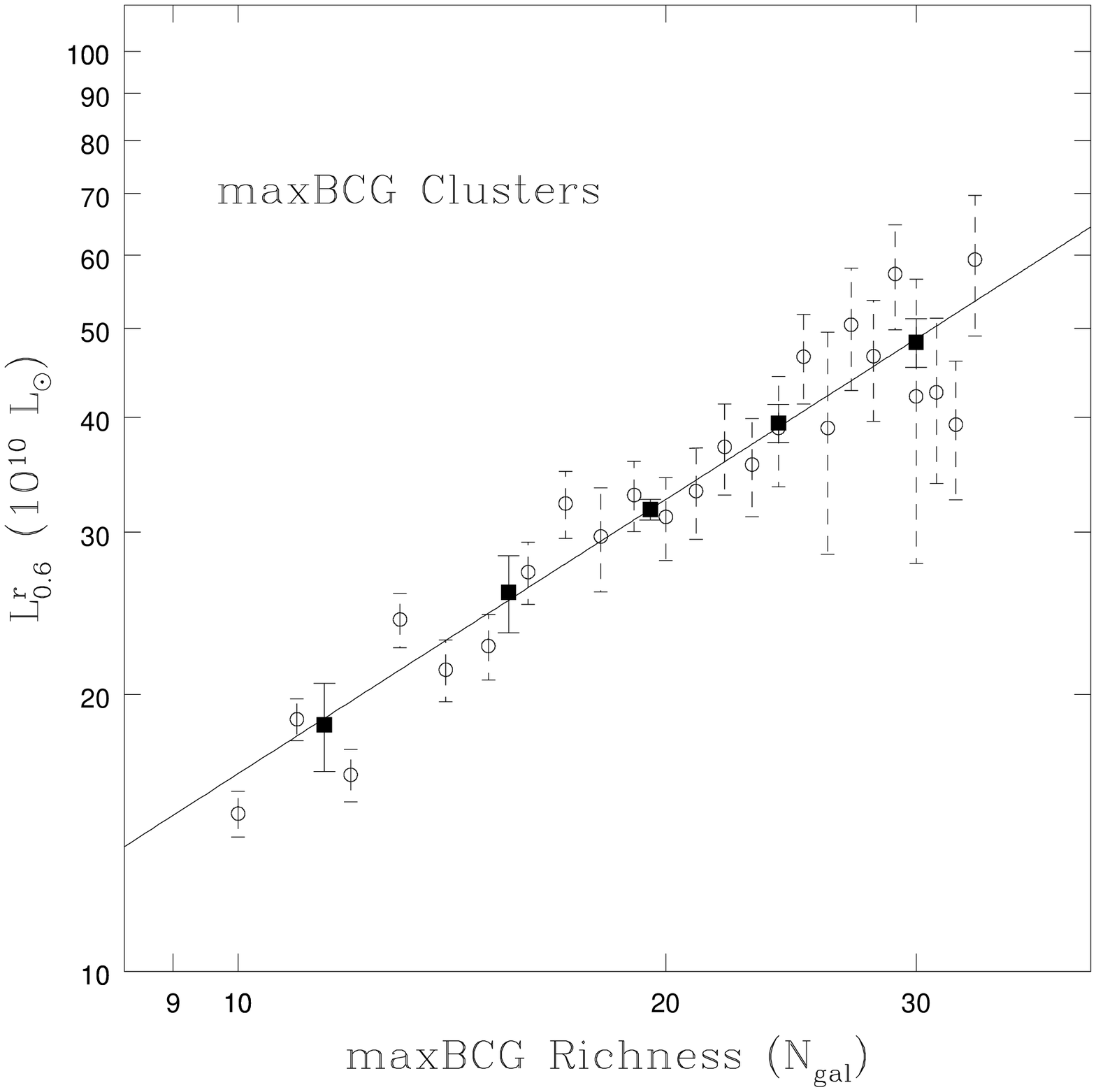}
\caption{Observed cluster luminosity versus richness for HMF and maxBCG clusters.
Cluster luminosity is observed in the $r$-band, within a radius of 0.6 $h^{-1}$ Mpc,
for stacked clusters at a given richness. The luminosities are k-corrected,
background subtracted, and integrated down to M$_r$ = -19.8. Dark squares 
represent binned data. (See section 3)
\label{f1}}
\end{figure}

\epsscale{1.0}
\begin{figure}
\plotone{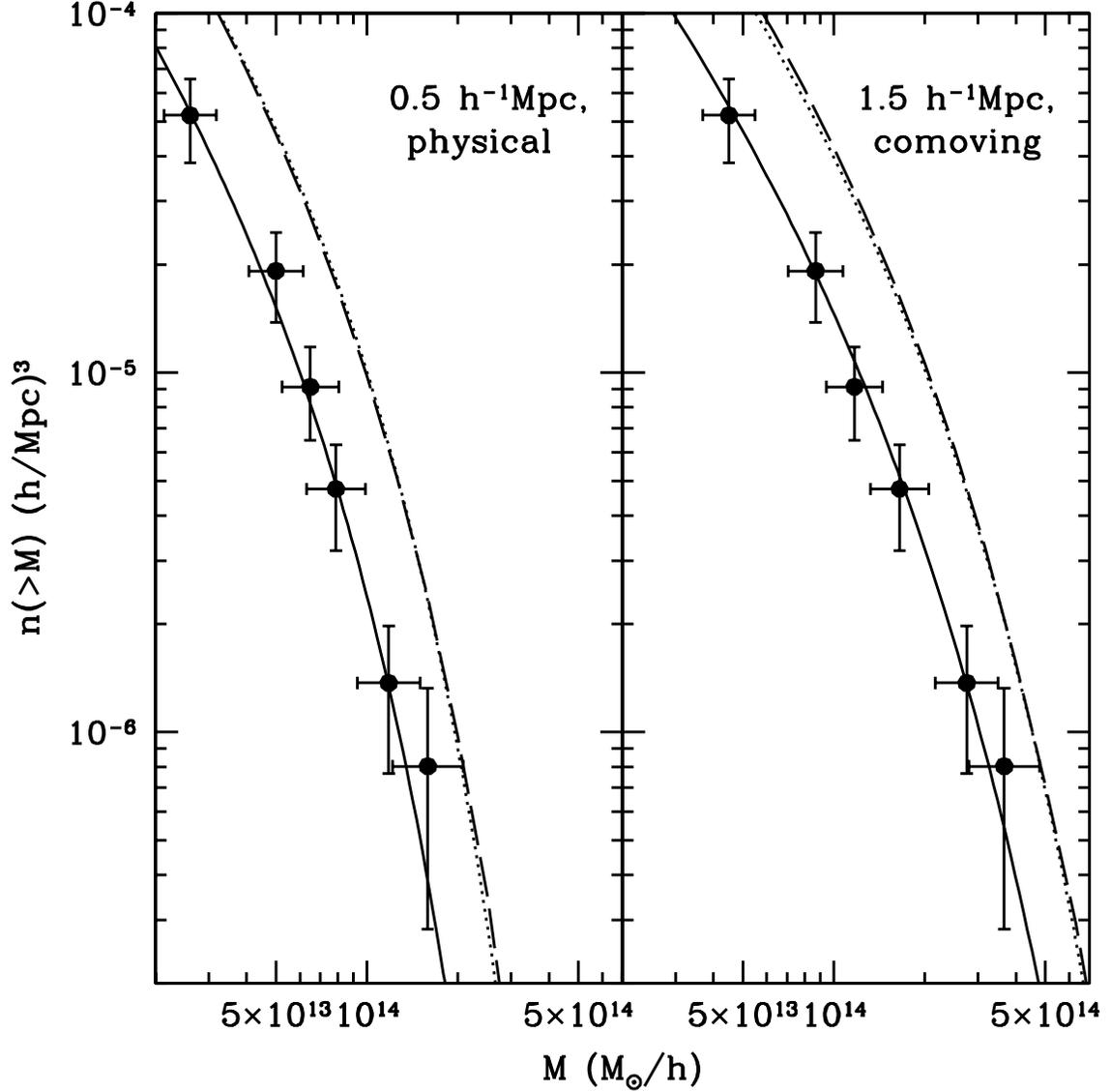}
\caption{The HMF cluster mass function for masses within 
radii of 0.5 $h^{-1}$ Mpc (left panel) and 1.5 $h^{-1}$ Mpc
comoving (right panel).  The solid line is the best-fit 
analytic mass function
(determined at 0.6$h^{-1}$ Mpc and extrapolated to the
appropriate radius assuming an NFW profile),  with \om =0.175 and
\sige =0.92.  In each panel the dashed line is the mass function
measured from an N-body simulation of the concordance LCDM model with \om =0.30 and \sige =0.90;  
the dotted line is the analytic prediction for this simulated cosmology.
\label{f2}}
\end{figure}

\begin{figure}
\plotone{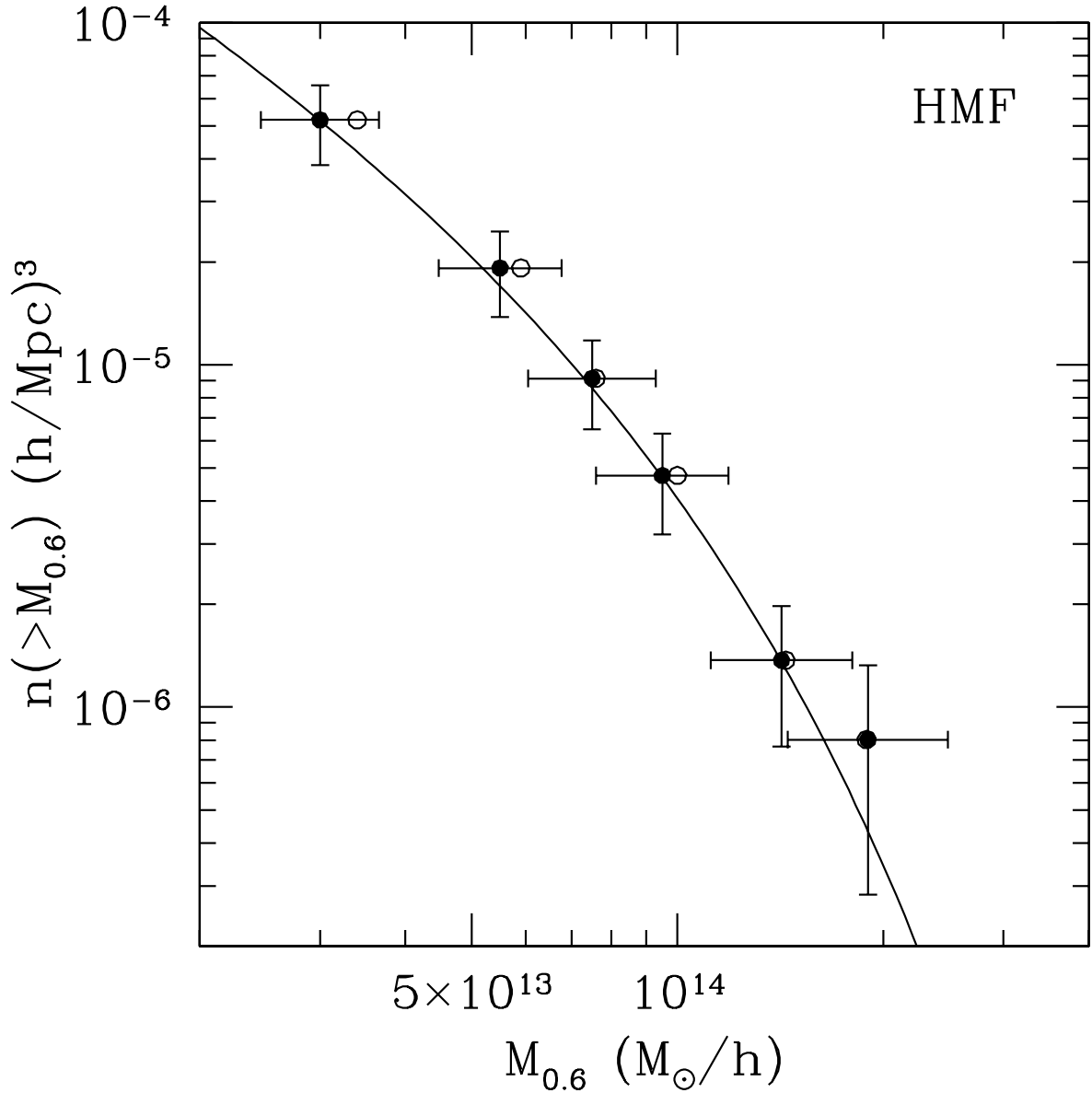}
\caption{The HMF cluster mass function,  showing masses (within 0.6 $h^{-1}$ Mpc) 
determined from both the luminosity -- mass calibration (filled circles) and the independent velocity 
dispersion -- mass relation (open circles).  (The observed cluster abundances assume a volume corresponding 
to a flat \om = 0.2 cosmology.)  
The best-fit analytic model,  with \om = 0.175 and \sige = 0.92,  is shown by the solid line.
\label{f3}}
\end{figure}

\epsscale{0.75}
\begin{figure}
\plotone{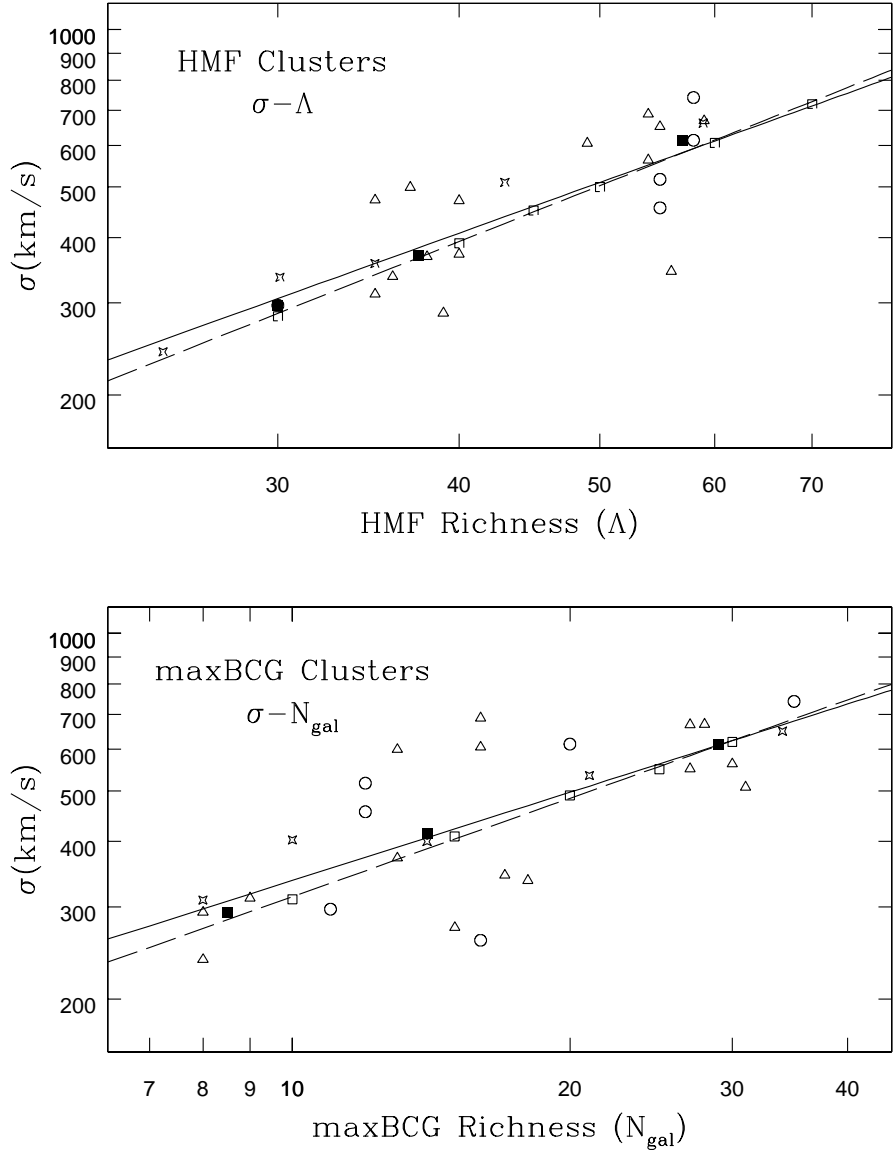}
\caption{Relation between observed cluster velocity dispersion $\sigma$ and cluster richness 
(triangles are SDSS observed velocity dispersions, circles are Abell clusters, dark squares are medians, and solid 
line is best fit to the velocity data. Stars represent SDSS observation of stacked cluster data, shown for comparison 
only). The median observed velocities are compared with the velocity determined from the cluster masses derived 
from the mean richness--luminosity--mass relation used in Section 3 (represented by open squares and the best-fit 
dashed line).  A good agreement between the two methods is seen. 
\label{f4}}
\end{figure}

\epsscale{1.0}
\begin{figure}
\plotone{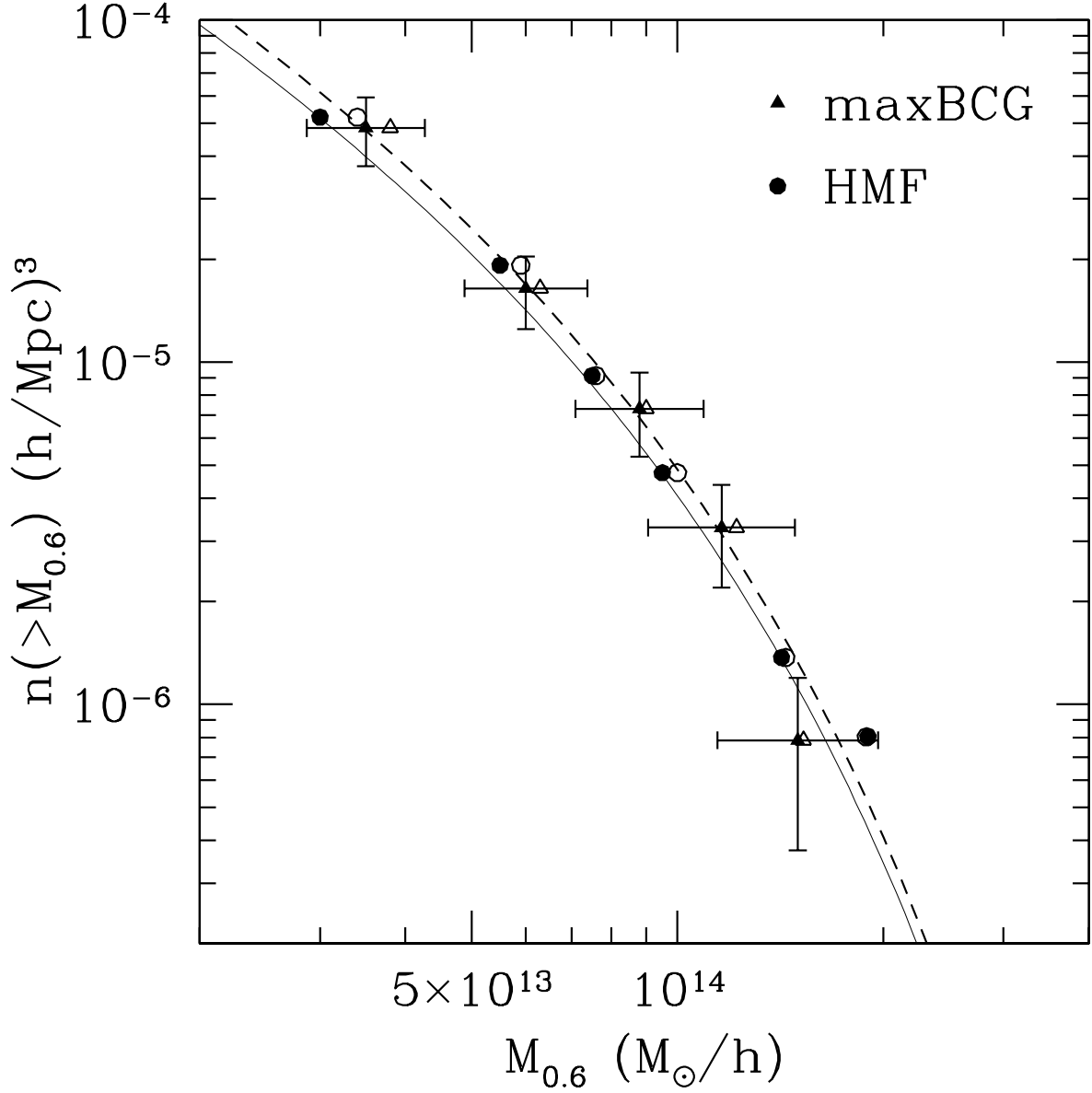}
\caption{The maxBCG and the HMF cluster mass functions,  showing masses determined from both 
luminosity -- mass relation (solid triangles: maxBCG; solid circles: HMF) and velocity dispersion -- 
mass relation (open triangles: maxBCG; open circles: HMF).  
The best-fit analytic models are shown by the dashed line 
(maxBCG; \om = 0.195, \sige = 0.90),  and solid line (HMF; \om = 0.175, \sige = 0.92;  as in Figure 2). 
\label{f5}}
\end{figure}

\begin{figure}
\plotone{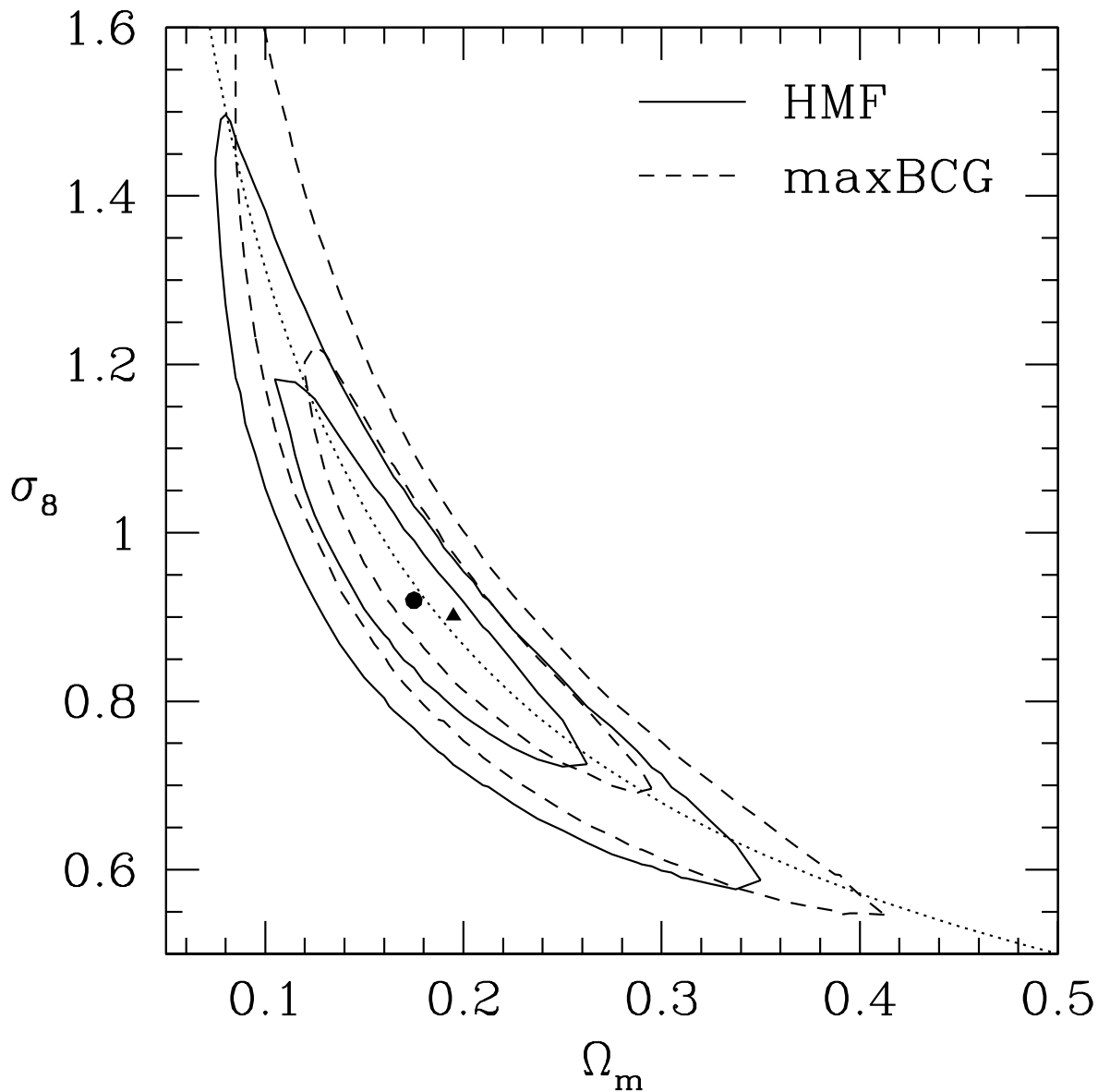}
\caption{Allowed \om --\sige \ range:  one-- and two--$\sigma$ confidence contours for
HMF (solid lines) and maxBCG clusters (dashed lines).  The dotted
curve is the best-fit relation $\sigma_8$ = 0.33$\Omega_m^{-0.6}$ $\simeq$ ($\frac{0.16}{\Omega_m}$)$^{0.6}$.  
The best-fit \om, \sige \ values are shown by the dark circle (HMF) and triangle (maxBCG).
\label{f6}}
\end{figure}

\epsscale{0.68}
\begin{figure}
\plotone{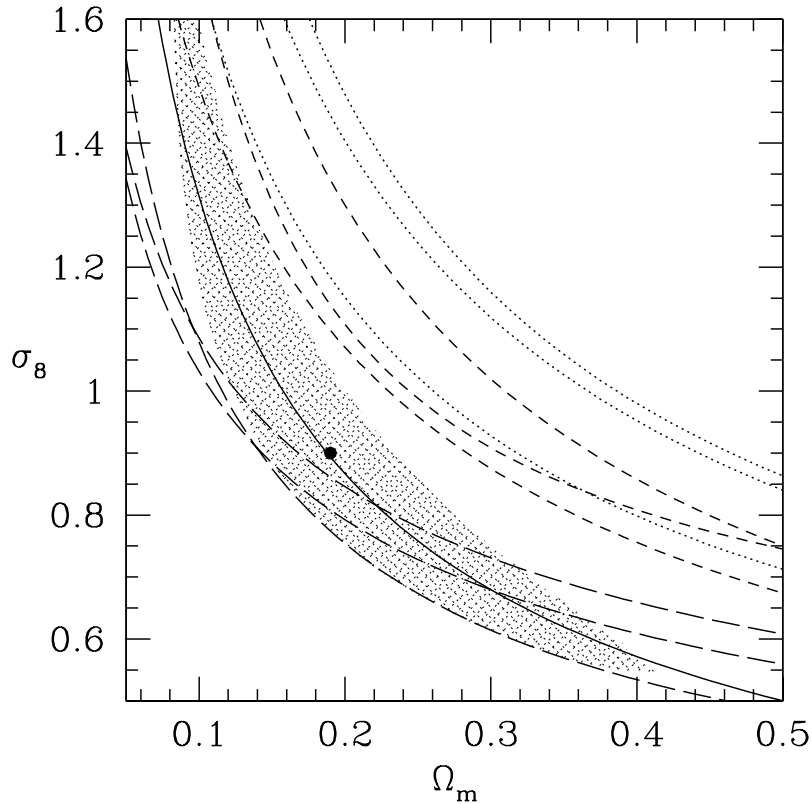}
\caption{Comparison with previous results: the allowed \om-\sige\  
range from our present SDSS mass function is compared 
with previous results from cluster temperature functions.
The current 2--$\sigma$ constraints (from Fig. 6) are represented by the shaded area, 
with the best-fit relation shown by the solid line, and the best-fit 
parameters (\om = 0.19, \sige = 0.9) indicated by the filled circle. The various dotted and dashed
curves represent the best-fit relations given by previous work (for clarity,
the allowed width of each range is not shown; it typically corresponds
to $\pm\sim$10$\%$ in \sige \ at a given \om). Dotted lines represent early results 
(\citealt{whi93, VL96, eke96}; top to bottom); 
short-dashed lines are mostly re-analyses of same/similar early data 
(\citealt{pie01, ouk01, pw01}; top to bottom); 
long-dashed lines represent lower normalization relations obtained from
recent analyses and/or recent samples (\citealt{sel02, rei02, via02}; top to bottom at higher \om). 
Weak lensing
analyses on large scales (Sections 1, 4; not shown) yield results that are 
mostly near the high range of these curves, with \citealt{hoe02} fitting
well on our best-fit relation. 
\label{f7}}
\end{figure}



\end{document}